# BUILDING AGILITY IN COVID-19 INFORMATION SYSTEMS RESPONSE IN SRI LANKA: RECOMMENDATIONS FOR PRACTICE

Pamod Amarakoon, University of Colombo, Sri Lanka, pamodm@gmail.com

Jørn Braa, University of Oslo, jbraa@ifi.uio.no

Sundeep Sahay, University of Oslo, sundeeps@ifi.uio.no

**Abstract:** COVID-19 pandemic tested the capacity of information systems in countries on the ability to rapidly respond to requirements which were not anticipated. This article analyzes the socio-technical determinants of agility in building the IS response to the COVID-19 pandemic in Sri Lanka. We deploy qualitative research methods to explore the case study of implementation of COVID-19 surveillance system in Sri Lanka. Three key recommendations are developed for practice relating to high-level multisectoral governance, use of lightweight digital platforms and leveraging on existing capacities and infrastructure.

## 1. INTRODUCTION

Pandemics are characterized by high levels of uncertainty as they go through unexpected and dynamically changing situations making their responses difficult to predict. What is certain, however, is that timely quality data is important for health services and policy makers to respond optimally. How to best develop information systems (IS) to respond to health crisis as pandemics is the topic of this article, based on an empirical analysis of the COVID-19 IS response in Sri Lanka.

When the H1N1 (swine flu) hit Japan in 2009, the software team applied agile iterative approaches because of the limitations of traditional 'waterfall' approaches in addressing rapid changes (Murota, Kato, & Okumura, 2010). Lai (Lai, 2018) emphasized the need to understand the determinants driving agility, and identified four such conditions: i) committed leadership, to help overcome organizational barriers; ii) enabling interdependence to allow coordination across multiple stakeholders; iii) context specific domain knowledge and expertise; and, iv) shared understanding of common goals shaping response.

Uncertainty in pandemics provides unique resources for innovative and swift action. In the context of COVID 19 response, Rigby, et. al. (Rigby, Elk, & Berez, 2020) argue that agile responses does not typically come from a strategic plan, but comes as someone, somewhere identifies an urgent need for action and finds innovative and non-standard solutions. Agile software development methods (Hamed & Abushama, 2013) is one such approach to build IS solutions based on rapid prototyping cycles of software development, release, feedback and improvements. However, arguably this approach is limited to the technical aspects, and there are other socio-technical conditions to consider (Lai, 2018). Given this backdrop, the research question this paper addresses is:

What are socio-technical determinants of agility in the context of IS response to COVID-19, and how can these be best materialized in practice?

We analyze this question is the context of an ongoing successful IS response effort to the pandemic in Sri Lanka. After this brief introduction, we conceptually discuss the notion of agility, followed by a discussion of the case and its analysis. Finally, some recommendations for practice are presented.

## 2. CONCEPTUAL FRAMING: SOCIO-TECHNICAL PERSPECTIVE ON AGILITY

Once WHO had declared the 2009 Swine flu a pandemic, Japan set up a task force to work on the IS response, while recognizing the 'situation caused by the unforeseen virus was beyond the architect's imagination' (Murota et al., 2010). The management decided to develop a new system (Day 1), and a prototype was rapidly developed following an "agile" approach (Day 4), and following feedback, a next version was developed (Day 8) and on Day 13, the system was ready for national roll-out. The rapid iterative cycles continued through the pandemic to accommodate to the different changes, such as in surveillance policies and pandemic phase transitions. For example, while the initial focus was on identifying all patients and suspects, it subsequently changed to locating cluster incidents with confirmed virus RNA. As the surveillance design was modified for each phase, the IS had to also rapidly evolve in support using agile methods (Lee & Xia, 2010) as traditional waterfall methods were grossly inadequate (Kapyaho & Kauppinen, 2015). Doz et al describes strategic agility as the ability of the organization to renew itself and remain flexible without sacrificing efficiency(Doz & Kosonen, 2008). From an organizational perspective to ensure strategic agility it require sensitivity to environment, resource fluidity and collective commitment.

While agile methods such as SCRUM emphasize individuals and interactions over processes and tools (Agile Alliance, 2001), we argue an important lack is the inadequate attention given to the materiality of the platform itself used for the agile development. We see "lightweight IT and platforms" as an important determinant of agility.

(Øvrelid & Bygstad, 2016) contrasts lightweight and heavyweight IT with respect to the knowledge regimes they underpin. Lightweight represents "a socio-technical knowledge regime" driven by competent users' need for solutions, enabled by the consumerization of digital technology, and realized through innovation processes. Usability, easiness in implementation and ability to quickly follow up pilots are described as key characteristics of lightweight IT. In contrast, heavyweight IT represents a knowledge regime driven by IT professionals, enabled by systematic specification and proven digital technology, and realized through software engineering processes. The focus on knowledge regimes highlights the socio-technical nature of IT.

We combine this notion with that of a platform, described as a "software-based product or service that serves as a foundation on which outside parties can build complementary products or services" (Tiwana, 2013). Tiwana also argues that platforms must be multisided, enabling the coming together of two or more actors such as end users and app developers. The manner in which a platform enables the coming together of different people, technologies and processes represents a platform architecture (Baldwin & Woodard, 2008), which consists of a core with a set of stable components, and complementary components that interact with the core. The wider software platform eco-system is therefore composed by software modules, apps, interfaces between them, developers, users and a variety of stakeholders. Henfridsson and Bygstad (Henfridsson & Bygstad, 2013) suggest the unit of analysis should not be the core of the platform but its boundary resources and the larger ecosystem. We use this perspective for analysing the determinants of agility.

## 3. METHODS

For this article, we used a case study approach because it helped us to look into the background and complexities of the phenomena – agile growth of HIS in a pandemic situation in an LMIC context like Sri Lanka. According to Darke et al, such a method is useful in younger and less developed research areas, such as the phenomenon studied in this context (Darke, Shanks, & Broadbent, 1998). Our methodology was an analytical inquiry exploring a contemporary phenomenon within a real-life framework – the Sri Lankan health system context, as described by Yin(Yin, 2014).

The basis of the development of the case study is an ongoing 10-year old engagement of the three authors in the strengthening of the national health information systems in Sri Lanka, including in the design and implementation of the COVID-19 IS response. The principal author is a graduate of Masters and Doctoral program of Health Informatics and a key implementer of the IS response to COVID pandemic in Ministry of Health (MoH) Sri Lanka. He conducted most of the field level data collection for the study while engaging in implementation activities. Two other authors are senior professors of the University of Oslo who have been closely engaged in establishment of Masters Program in Biomedical Informatics in the University of Colombo, Sri Lanka and have engaged in academic program of the Masters and Doctoral programs as well as providing mentorship to health informatics implementations in Sri Lanka. For this paper, we focus firstly on the issue of building national health informatics capacity, which provides the context for the analysis of the IS response to COVID-19.

As argued by Yin we used multiple data collection techniques (Yin, 2014). We conducted 10 in-depth interviews from stakeholders at national and district level engaged with the information management on COVID-19. These included 3 health informatics experts from MoH, 2 health administrators from national level, 2 experts in ICT outside the MoH who contributed to the surveillance system, 3 implementers and trainers of the system and 2 district level implementers of the surveillance system. Since some of the writers were closely involved in the planning, creation, and execution of the HIS in question, their lived experiences were recorded in the form of narratives and analyzed. Memos were also compiled from observations made at stakeholder meetings during the setup of the DHIS2-based surveillance scheme. Secondary data was also included in this analysis, including publications released by public health organizations, records delivered at stakeholder meetings, and observations taken at informal meetings with stakeholders. During the study, the authors' own historical accounts of the Masters program's establishment, the graduates' contributions to the digital health ecosystem, and the development of health informatics in Sri Lanka were also taken into account. However, since this was a retrospective analysis, the data collection for critical stakeholder meetings had to depend on the authors' own recollections and meeting minutes.

Throughout the data analysis process, we were continuously sharing our thoughts and perspectives on the information we had gathered. The case narrative was then collaboratively developed, and thematic analysis methodology was used to define evolving themes, which were then critiqued before being used in the development of this paper, as proposed by Braun et al (Braun & Clarke, 2014). We were able to eliminate any possible bias that could have clashed with our understanding of the case by taking a collaborative approach to data analysis, considering the responsibilities we perform within the phenomena we wish to examine.

## 4. CASE STUDY

### 4.1. Building of national health informatics capacity

This process of building health informatics capacity in Sri Lanka started in 2009, and it was of immense significance in building the COVID-19 response.

Sri Lanka provides both health and education free to its citizens, including the highly sought-after medical education. This policy has inspired more than 90% of the medical graduates join the government medical services. Furthermore, the Ministry of Health (MoH) also provides full funding for pursuing postgraduate specialization studies, including in health informatics, at the Postgraduate Institute of Medicine of University of Colombo. In 2009, the Masters in Biomedical Informatics (BMI) degree programme was established with financial and technical support from the University of Oslo, Norway. This programme had a futuristic vision of producing medical specialists with a background of biomedical informatics, so as to develop national sustainable capacity. The collaboration with Oslo and their world leading Health Information Systems Programme (HISP) has helped the BMI students to be educated in modern informatics concepts and build expertise on the DHIS2 platform (see dhis2.org). The unique action research orientation of HISP, was also an underlying pedagogical principle in the BMI programme and students-built expertise on how to apply DHIS2 to different problem contexts, such as for malaria and TB information management.

On completion of the programme, the BMI graduates reverted to the MoH where they assumed duties as 'Medical Officers in Health Informatics' responsible for aspects of health policy, system design, development, implementation and training. These officers, with rare hybrid skills of medicine and informatics, become crucial technical and administrative experts driving the national HIS forward, and were fundamental in building the COVID-19 IS response, now described.

**4.2. Design and implementation of the COVID-19 IS response**

From January 2020, when there was a rising concern of COVID-19 cases in some East Asian countries, Sri Lanka recognized the threat since it is a country of high tourist traffic and also hosts many Chinese based infrastructure projects. They acted rapidly, setting up a Presidential task force headed by the commander in chief of the army, with key roles assigned to the Minister of Health, Director General of Health Services (DGHS) and the police. The authority wielded by the task force members ensured rapid decisions could be taken on both the health and citizen-based issues, such as on quarantine and contact tracing. The task force reached out to representatives from other sectors such as agriculture, education, immigration and social services, and sub-committees created on implementation responsibilities.

An urgent priority of the task force was to establish a robust IS response to gather information on possible sources of infection and follow up of suspect cases. Immediate bottlenecks was the unavailability of a generic outbreak management IS in the MoH and the long bureaucratic process of procurement. The task force had the authority to subvert the typical bureaucratic processes and decided to adopt a customisable platform as opposed to developing something from scratch. Given the need to share data across departments, a cloud-based solution was mandated.

The MoH selected the DHIS2 platform for building this system, as it met identified criteria and could quickly be configured for use, and because there was existing capacity and experiences with the system. DHIS2 and its developers (HISP Sri Lanka) had the prior trust of the government. Initial discussions between the ministry and HISP Sri Lanka started in 3rd week of January, which identified the initial priority of building a port of entry tracker module. Using the existing DHIS2 "tracker" functionalities, this module was rapidly discussed and presented to the DGHS on 27th January just prior to the country reporting its first case of coronavirus. Permission was obtained for national roll out.

Sri Lanka reported its second COVID-19 case on 11th of March, which was also accompanied with large numbers of nationals returning from high burden countries such as Italy and South Korea, leading to a steady rise in cases, going over 100 in 2 weeks. Confirmed cases were admitted to designated hospitals and tracking of contacts was initiated. The country closed the ports of entry in the third week of March, and the IS response shifted from port of entry to tracking of confirmed and suspect cases from community clinics, quarantine centres and testing laboratories. DHIS2 is a generic platform and a new module for tracking persons during their quarantine period was rapidly developed.

The country reported its first COVID-19 death on 28th of March, and the current total is at 11 cases, including 7 cases in the first 11 days. This surge and increase in number of patients requiring the ICU care were seen as red flags, requiring optimal management of ICU beds. However, the country's critical care IS had not been operational in the last few years, and so the task force required an ICU bed management component to be included to detect availability and proximity of ICU beds. Following this request, a module in DHIS2 to track availability of ICU beds was developed during a hackaton with participants from local software development and volunteer community and with input from healthcare professionals.

Typical Ministry's capacity building initiatives were based on face-to-face settings which was now not possible. After the declaration of the national curfew on 21st of March, training programmes were mobilized through the Zoom videoconferencing platform, and training content was developed with high graphical content rather than having lengthy training material. This shift in training media and content enabled the Ministry to rapidly scale the system nationally. The training sessions were typically targeted for each district with participation of around 20 users per online training session.

While the overall IS response reflected well the process of local innovations, there were also innovations developed through collaboration with the Oslo global DHIS2 team. Some new requirements could not be addressed locally and required work on the core software, which could be done with the global team. For example, while developing the port of entry module, the ministry required the flexibility to change the registering organization unit of a passenger, a functionality not available in the core DHIS2. The global team then helped modify the codebase and to also help visualize transmission relationship and contact mapping as a graphical network. This example shows the power of global-local collaboration and sharing of expertise. Slowly, as the system started to acquire more generic features, the global team worked towards creating a generic app which is now in use in 20+ countries, including Norway. Additionally, local software development capacity was organized through multisector collaboration between MoH, government ICT agency and the volunteer software developer. A hackathon was organized in the second week of March to develop additional modules and functionalities, such as the ICU bed tracking module mentioned above. The ICT Agency in addition provided infrastructure support for hosting of the solution in the government cloud, internet facilities to hospitals and health facilities as well as provision of workspace environments for collaborations such as "Slack". Thus, there were multisector engagements across domains at local and international level.

We summarize the above discussion through a table which summarizes the different functionalities/modules developed and implemented using the DHIS2 platform. The speed and variety of development is illustrating the agility involved.

Table 1: Modules of the COVID-19 Surveillance System and Implementation Process

| Module and date of release | Implementation process around lightweight DHIS2 platform |
|---|---|
| Port of Entry Module<br>27th January 2020 | • Rapid approval by task force<br>• DHIS2 inbuilt functionalities allowed for rapid customization by local experts and presentation to DGHIS<br>• Rapid end user training |
| Hackathon & Multisector Discussions<br>14th March 2020 | • Strengthening multi-sectorial collaboration<br>• Building and extending existing Web API and Web Apps platforms<br>• Global expertise supporting local innovation |
| Development of Case & Suspect Monitoring Module<br>15th March 2020 | • Use existing DHIS2 tracker functionality used to quickly customize the module<br>• Customization by super users, without core programming expertise<br>• Rapid end user training |
| Contact Mapping Application & Mobile Tower Location Integration<br>23rd March 2020 | • New functionality developed by extending existing platform.<br>• Globalization of local innovations through HISP network |
| Quarantine Persons Module<br>25th March 2020 | • Use existing DHIS2 tracker functionality used to quickly customize the module |
| | • Customization by super users, without core programming expertise<br>• Rapid end user training |
| MyHealth Citizen Information Mobile App<br>27th March 2020 | • Support integration |
| Health Sector Resource Monitoring Module<br>5th April 2020 | • Customization by super users, without core programming expertise<br>• Rapid end user training |
| ICU Bed Management Web App<br>11th April 2020 | • Building and extending existing Web API and Web Apps platforms |
| Laboratory Reporting Module<br>25th April 2020 | • Customization by super users, without core programming expertise<br>• Rapid end user training |
| Generic Relation Tracing Web App<br>10th May 2020 | • Globalization of local innovations |
| Scaling-up implementation<br>May – October 2020 | • Training of users and expansion |
| More focus on laboratory data collection and custom app for laboratory data management<br>November – December 2020 | • Involvement of volunteer developers for laboratory data management app |
| Vaccination Tracker<br>January – March 2021 | • Collaboration of WHO country office to design COVID Vaccine Tracker<br>• Shared metadata with University of Oslo and DHIS2 community<br>• First country to deploy COVID vaccine tracker<br>• Incorporation of digital certificate |

## 5. DISCUSSION

Our case analysis identifies three key determinants of agility which we first present through the model below and then discuss. In the section that follows, we present recommendations for practice based on the model (Figure1).

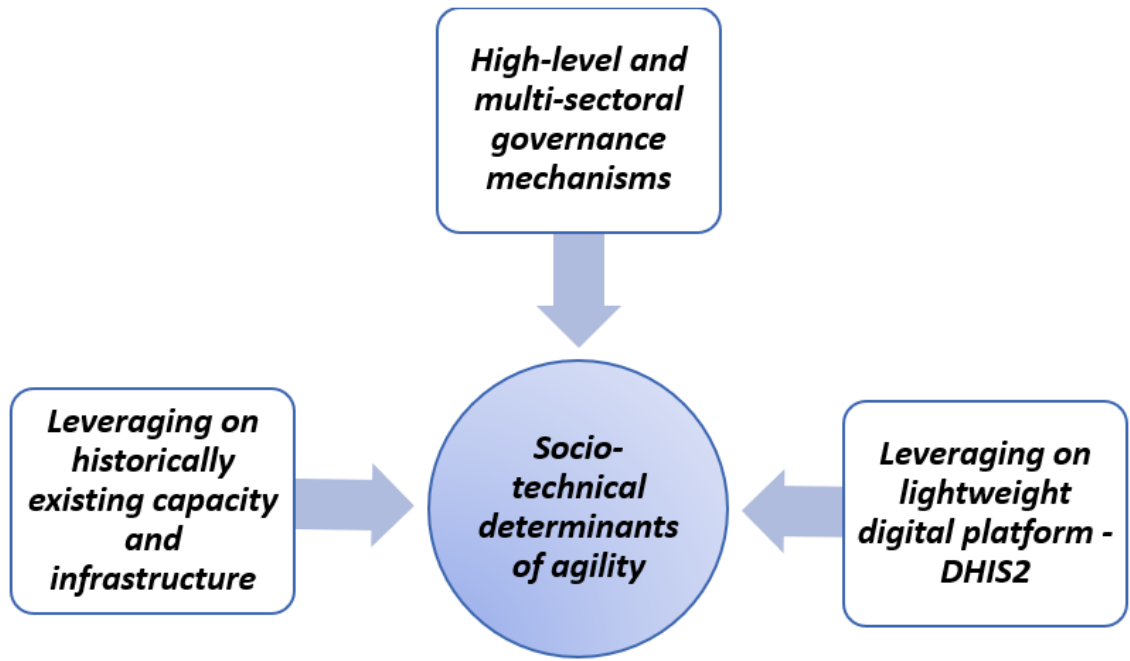

Figure 1: Socio-technical determinants of agility in IS response

### 5.1. Socio-technical Determinants of Agility in the Times of Uncertainty

Our model emphasizes that agility is not only about software development methods, but involves a combination of social, political and technical conditions, including: i) high-level and multi-sectoral governance mechanisms; ii) leveraging on historically existing capacity and infrastructure; iii) use of established lightweight digital platform.

**High-level and multi-sectoral governance mechanisms**: Even before the pandemic had made its entry in the country, a high-level task force was constituted under the leadership of the President's office and comprising of senior decision makers from different sectors. This governance structure enabled the taking of agile decisions, which were cross-departmental and broke down historically existing bureaucratic structures, such as related to the software procurement process. Similarly, questions of quarantine and lockdowns were decided by representatives from health, immigration and police. The structure enabled multi-sectorial rapid decision-making, with well thought through processes to ensure implementation compliance.

**Leveraging on historically existing capacity and infrastructure**: Three aspects of history were crucial in building agility. One, concerned the BMI programme which had been in operation since 2009 and had contributed to the education of many doctors who occupied important health informatics positions in the Ministry. Two, was the fact that the DHIS2 was well established through use by nearly 20 health departments. Three, was the HISP Sri Lanka team, comprised of graduates from the BMI programme with high levels of DHIS2 expertise and strong collaborative links with the global DHIS2 team to bring about changes in the core when needed.

**Leveraging on lightweight digital platform – DHIS2**: The DHIS2 was a well-established digital platform within the Ministry, and over the years had cultivated an enabling infrastructure around it, such as for server hosting. The DHIS2, by design, is a lightweight platform that is flexible and adaptable to different use contexts, and new functionalities can be added relatively easily, without the need for core programming support. While as an web based independent application DHIS2 was easy to implement within the existing infrastructure, integration with other systems such as labs and custom was also easily achieved. The DHIS2 comes with core modules, for example for data analysis, dash boards, spatial analysis etc, which were easily adapted to the specific requirements for COVID-19 surveillance. For features, not currently available in the core, for example related to conducting network analysis, could be developed by enrolling national and global expertise, for example through hackathons.

In summary, agility was ensured through: i) the high-level governance mechanism, which enabled rapid decision making and implementation; ii) the lightweight platform allowed for rapid incorporation of new functionalities as the information demands of the pandemic evolved; and, iii) the existing DHIS2 related capacity which allowed system development and implementation to take place on the fly, without needing to build from scratch..

### 5.2. Recommendations for practice

**Establishing high level governance mechanisms**

Many existing problems of HIS in LMICs relate to the absence of effective governance mechanisms to mitigate the technical and institutional fragmentation of systems (Sahay, Sundararaman, & Braa, 2017). This leads to challenges in establishing and implementing standards, promoting fragmentation and the absence of an architecture approach. Pandemic responses by design require cross-sectoral collaboration across ministries of health, finance, home and others, which has been historically difficult to achieve in practice. Our case illustrates, a high-level task force in conditions of crisis can help to ensure such collaboration. In different countries, the models of building such a structure would be different, for example housed in the President's office (as in this case), or in other departments such as home or health, depending on the political power they wield and their capacity to engage in multi-sectorial collaborative action.

**Policy promoting the use of lightweight digital platforms**

While there is an implicit acknowledgement of the value of free and open source digital platforms, there is an absence of explicit policy on its implementation. Given the need for sharing data across systems, there is an urgent need for using free software which do not come with licensing encumbrances and have open APIs which are publicly published. For example, in the case presented, the use of free software allowed easy sharing of data between the immigration and health departments, which would have been impossible in its absence. The lightweight aspect of platforms ensures that they can be developed and evolved by super users without core software programming expertise.

**Focus on continuous and institutional capacity building**

New digital platforms require the development of novel multi-faceted capacities, which take time and effort to build. These capacities cannot be developed through individual training programmes, as is often the case in development projects, but need to be anchored in institutions. The BMI programme at the University of Colombo provided that anchoring and their symbiotic relationship with the Ministry ensured hybrid expertise was developed and rooted in the country. This solid foundation allowed for new capacities to be quickly developed around the new application, and using innovations in training media and content, scale could be achieved. An important policy implication is to develop a strong foundation of capacity anchored in institutions with symbiotic relationships with the Ministry.

## 6. CONCLUSION

Pandemics by definition are dynamic and evolving rapidly requiring also the IS response to be flexible and evolve in an agile manner. Understanding how such agility can be enabled is thus both a crucial theoretical and practical challenge. Our paper has sought to contribute to building understanding on how to build agility, emphasizing it is not a technical but fundamentally a socio-technical collaborative effort.